\documentclass{PoS}

\usepackage{amsmath}
\usepackage{amssymb}
\usepackage{subfigure}

\newcommand{\g}[1]{\gamma_{#1}} 
\renewcommand{\l}{\left}
\renewcommand{\r}{\right}
\newcommand{\tr}{\mathrm{tr}}

\newcommand{\RM}{\mathbb{R}_M}
\newcommand{\expect}[1]{\left\langle #1 \right\rangle}
\newcommand{\stat}{\mathrm{stat}}
\newcommand{\sys}{\mathrm{sys}}

\title{
\vspace{-3.0cm} 
{\small \normalfont \hfill DESY 15-204 \\} 
\vspace{2.4cm} 
Testing the Witten-Veneziano Formula on the Lattice}

\ShortTitle{Testing the Witten-Veneziano Formula on the Lattice}

\author{Krzysztof Cichy$^{a,b,c}$, Elena Garcia-Ramos$^{b,d}$, Karl Jansen$^{b}$, \speaker{Konstantin Ottnad}$^{e}$, Carsten Urbach$^{e}$ \\\par
        $^a$ Goethe-Universit\"at, Institut f\"ur Theoretische Physik, Max-von-Laue-Stra\ss e 1, D-60438 \\ 
        $\ \,$ Frankfurt a.M., Germany \\
        $^b$ NIC, DESY, Platanenallee 6, D-15738 Zeuthen, Germany \\
        $^c$ Adam Mickiewicz University, Faculty of Physics, Umultowska 85, 61-614 Poznan, Poland \\
        $^d$ Humboldt Universit\"at zu Berlin, Newtonstr. 15, D-12489 Berlin, Germany \\
        $^e$ Institut f\"ur Strahlen- und Kernphysik (Theorie), Nussallee 14-16 and Bethe Center for \\
        $\ \,$ Theoretical Physics, Nussallee 12, Universit\"at Bonn, D-53115 Bonn, Germany \\ \par
        E-mail: \email{krzysztof.cichy@desy.de}, \email{elenagr@ifh.de}, \email{Karl.Jansen@desy.de}, \email{ottnad@hiskp.uni-bonn.de}, \email{urbach@hiskp.uni-bonn.de}}

\abstract{In this proceeding contribution we report on a test of the famous Witten-Veneziano formula using lattice techniques. We perform dedicated quenched simulations and apply the spectral projector method to determine the topological susceptibility in pure Yang-Mills theory. In order to compute the relevant meson masses and the flavor singlet decay constant we employ lattice QCD with $N_f=2+1+1$ dynamical Wilson twisted mass fermions. Taking the continuum and the SU$(2)$ chiral limits we find good agreement within uncertainties for both sides of the formula.}

\FullConference{The 33rd International Symposium on Lattice Field Theory\\
                14 -18 July 2015\\
                Kobe International Conference Center, Kobe, Japan}

\graphicspath{{}}

\begin{document}
\section{Introduction} 
The Witten-Veneziano formula \cite{Witten:1979vv,Veneziano:1979ec} in the chiral limit
\begin{equation}
 \label{eq:W-Vformula_chiral}
 \mathring{M}_{\eta^{\prime}}^2=\frac{4N_f}{f_0^2}\chi_{\infty}\,.
\end{equation}
relates the mass $\mathring{M}_{\eta^{\prime}}$ of the $\eta'$ in the chiral limit to the topological susceptibility in pure Yang-Mills theory $\chi_\infty$ and the flavor-singlet decay constant $f_0$. Therefore, it provides an explanation for the large mass of the $\eta'$ compared to the octet mesons through non-trivial topological fluctuations of the gauge fields. In the presence of non-vanishing quark masses Eq.~(\ref{eq:W-Vformula_chiral}) reads 
\begin{equation}
 \label{eq:W-Vformula}
 \frac{f_0^2}{4N_f}(M_\eta^2+M_{\eta'}^2-2M_K^2)=\chi_{\infty} \,.
\end{equation}
where we have reshuffled terms compared to Eq.~(\ref{eq:W-Vformula_chiral}) to isolate $\chi_\infty$ on the r.h.s. of the equation. We remark that from a modern point of view the Witten-Veneziano formula represents a leading order result in chiral perturbation theory using a combined power counting scheme in quark masses $m_q$, momenta $p$ and $1/N_c$ \cite{Feldmann:1999uf}.

In this study we employ lattice methods to test the Witten-Veneziano formula directly. To this end we compute both the topological susceptibility in pure Yang-Mills theory (or quenched QCD) and the meson masses and the singlet decay constant $f_0$ in full QCD. While in this proceeding contribution we give an overview of our results with focus on the computation of $\chi_\infty$ and $f_0$, we refer to the recent publication \cite{Cichy:2015jra} for additional details.

\section{Quenched $\chi_\infty$}
For the computation of $\chi_\infty$ in pure Yang-Mills theory we performed dedicated simulations at four different lattice spacings in the quenched setup using the Iwasaki \cite{Iwasaki:1985we} gauge action
\begin{equation}
 S_G[U]=\frac{\beta}{3}\sum_x\left(b_0\sum_{\genfrac{}{}{0pt}{}{\mu,\nu=1}{1\leq\mu<\nu}}^4\mathrm{Re} \tr \left(1-P^{1\times 1}_{x;\mu\nu}\right)+b_1\sum_{\genfrac{}{}{0pt}{}{\mu,\nu=1}{\mu \neq \nu}}^4\mathrm{Re} \tr \left(1-P^{1\times 2}_{x;\mu\nu}\right)\right)\,,
 \label{eq:iwasaki}
\end{equation}
where $b_1=-0.331$ and $b_0=1-8b_1$. To perform these simulations we used the HMC algorithm implemented in the tmLQCD package \cite{Jansen:2009xp}. The details of the quenched simulations are compiled in Tab.~\ref{tab:qchddetails}, including the critical values $\kappa_c$ required to obtain $\mathcal{O}(a)$ improvement which were computed following the strategy introduced in \cite{Jansen:2005gf}. \par

\begin{table}[t!]
 \centering
 \begin{tabular}{@{\extracolsep{\fill}}cccccccc}
  \hline\hline
  $\beta$ & $T/a \times (L/a)^3$ & $r_0/a$ & $a$ [fm] & $a\mu$ & $\kappa_c^{\chi}$ & $Z_S/Z_P$ \\
  \hline\hline
  2.37 & $40\times20^3$ & 3.59(2)(3) & 0.1393(14) & 0.0087 & 0.158738 & 0.680(1)(27) \\
  2.48 & $48\times24^3$ & 4.28(1)(5) & 0.1182(14) & 0.0073 & 0.154928 & 0.707(1)(19) \\
  2.67 & $64\times32^3$ & 5.69(2)(3) & 0.0879(6)  & 0.0055 & 0.150269 & 0.752(1)(7)  \\
  2.85 & $80\times40^3$ & 7.29(7)(1) & 0.0686(7)  & 0.0043 & 0.147180 & 0.787(1)(3)  \\
  \hline\hline
 \end{tabular}
 \caption{Input parameters of the pure gauge ensembles and results for $Z_S/Z_P$. The errors of $r_0/a$ correspond to statistical and systematic uncertainties, respectively. The first error for $Z_P/Z_S$ is statistical, while the second one is systematic, accounting for the residual dependence on the mass parameter $M$.}
 \label{tab:qchddetails}
\end{table}

In order to compute $\chi_\infty$, we use the method of spectral projectors \cite{LuscherGiusti,Luscher:2010ik,Cichy:2013rra}. The definition of the topological susceptibility in terms of the spectral projector $\RM$ is given by 
\begin{equation}
 \label{eq:topsusspproj}
 \chi_\infty=\frac{Z_S^2}{Z_P^2}\frac{1}{V}\expect{\tr\{\gamma_5 \RM^2\}\tr\{\gamma_5 \RM^2\}}\,,
\end{equation}
and we refer to the original papers for further details about the method \cite{LuscherGiusti,Luscher:2010ik}. The spectral projector $\RM$ is defined as an orthogonal projector to the subspace of fermion fields spanned by the eigenvectors of the massive Hermitian Dirac operator $D^\dagger D$ corresponding to eigenvalues below some value $M^2$. To achieve $\mathcal{O}(a^2)$ scaling towards the continuum limit, the renormalized value of the threshold $M$, denoted by $M_R$, has to be fixed for all ensembles. In principle, the value of $M_R$ is arbitrary, however, one should avoid small values close to the renormalized quark mass and impose $aM_R\ll1$ to prevent enhanced cut-off effects \cite{LuscherGiusti}.

Furthermore, we remark that the ratio of pseudoscalar and scalar flavor non-singlet renormalization constants $Z_P/Z_S$ in Eq.~(\ref{eq:topsusspproj}) differs from unity for Wilson-type fermions without exact chiral symmetry. To compute $Z_P/Z_S$ for our quenched ensembles, we again use spectral projectors, following an approach which was first proposed in Ref.~\cite{LuscherGiusti}. The results are detailed in Tab.~\ref{tab:qchddetails}.

For a test of the Witten-Veneziano formula, one needs to take the continuum limit for $\chi_\infty$. Therefore, we have matched the physical situation for the quenched ensembles to the dynamical simulations. The first ensemble at $\beta=2.67$ was generated with a value of $r_0/a$, corresponding to the one obtained in the dynamical simulations for $\beta=1.95$. The physical volume and the value of $a\mu=0.0055$ used for the spectral projector method were matched to the corresponding values of one of the dynamical ensemble ($B55.32$ in the notation that was introduced in Ref.~\cite{Baron:2010bv} for labeling the ensembles). For further lattice spacings we kept $r_0\mu$ invariant and the physical volume constant at $L\approx2.8~$fm (assuming $r_0=0.5~$fm). 

\begin{figure}[t]
 \begin{center}
  \includegraphics[width=0.5\textwidth]{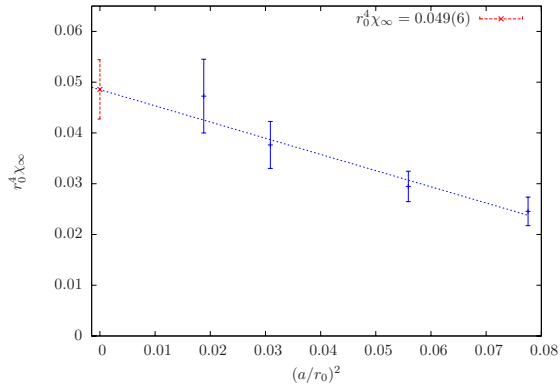}
 \end{center}
 \vspace{-0.75cm}
 \caption{Continuum limit extrapolation of $\chi_\infty$ as a function of $(a/r_0)^2$ for the quenched ensembles.}
 \label{fig:quenched}
\end{figure}

As discussed in Ref.~\cite{Cichy:2014yca}, our definition of $\chi_\infty$ is in principle affected by short distance singularities. At maximal twist, however, all terms that are linear in $a$ vanish and $\chi_\infty$ is $\mathcal{O}(a)$ improved, which leads to the expectation of $\mathcal{O}(a^2)$ scaling towards the continuum limit. This behavior is indeed confirmed by our data as shown in Fig.~\ref{fig:quenched}, hence we have performed a linear extrapolation in $(a/r_0)^2$ to extract our final, continuum result
\begin{equation}
 r_0^4\chi_\infty= 0.049(6)_{\rm stat+sys}\,,
 \label{eq:quenched_result}
\end{equation}
where the error includes the statistical and systematic errors in combined quadrature.

\section{Dynamical simulations}
For the calculation of the l.h.s. of Eq.~(\ref{eq:W-Vformula}) we use 16 ensembles with $N_f=2+1+1$ dynamic quark flavors provided by the European Twisted Mass Collaboration (ETMC) \cite{Frezzotti:2000nk, Baron:2010bv}. The ensembles used in this study cover light quark mass range corresponding to pion masses from roughly $220~$MeV to $490~$MeV, three lattice spacings ($A$, $B$ and $D$--ensembles in the notation of \cite{Baron:2010bv}) as well as different physical volumes. For a detailed list of the ensembles we refer to the published version in Ref.~\cite{Cichy:2015jra}. The required flavor-singlet masses $M_\eta$, $M_\eta'$ as well as $M_\mathrm{PS}$, $M_K$ and the decay constants $f_\mathrm{PS}$, $f_K$, have been computed in previous studies \cite{Baron:2010bv,Ottnad:2012fv, Michael:2013gka}. The flavor non-singlet decay constants do not appear in the Witten-Veneziano formula itself, but they are required for our computation of $f_0$.

\subsection{$\eta$,$\eta'$--mixing and $f_0$}
In general, decay constants are defined from axial vector matrix elements
\begin{equation}
 \l< 0 \r| A_\mu^a\l(0\r) \l| P\l(p\r) \r> = i f^a_P p_\mu \,,
 \label{eq:matrix_element_A}
\end{equation}
where $A_\mu^a\l(0\r)$ denotes the axial vector current with flavor structure denoted by the index $a$. In the octet-singlet basis the most general mixing scheme is given by
\begin{equation}
 \l( \begin{array}{ll}
   f_\eta^8 & f_\eta^0 \\
   f_{\eta'}^8 & f_{\eta'}^0 
  \end{array}\r) = \l( \begin{array}{rr}
   f_8 \cos \phi_8  & -f_0 \sin \phi_0 \\
   f_8 \sin \phi_8 & f_0 \cos \phi_0
  \end{array} \r) \,,
  \label{eq:singlet_octet_basis_parametrization}
\end{equation}
where we assumed exact isospin symmetry and neglecting possible contributions from the charm quark and mixing with further states such as glueballs. However, on the lattice we prefer to work in the quark flavor basis, where $A_\mu^0$ and $A_\mu^8$ are replaced by
\begin{equation}
 A^l_\mu = \frac{1}{\sqrt{2}} \l(\bar{u} \g{\mu} \g{5} u + \bar{d} \g{\mu} \g{5} d\r) \,, \quad A^s_\mu = \bar{s} \g{\mu} \g{5} s \,,
\end{equation}
leading to a new mixing scheme which exhibits different parameters ($f_{l,s}$, $\phi_{l,s}$) but still has the same form as Eq.~(\ref{eq:singlet_octet_basis_parametrization}). Furthermore, it turns out that we cannot directly use axial vector operators, because the resulting signals turn out to be too noisy, which is why we resort to pseudoscalar operators. However, it is possible to relate the two schemes as well as axial vector and pseudoscalar amplitudes through chiral perturbation theory ($\chi$PT), which ultimately allows to obtain expressions for $f_0$ in terms of quantities that can be computed on the lattice
\begin{align}
 f_0^2 &= - 7/6 f_{\rm PS}^2 + 2/3 f_K^2 + 3/2 f_l^2 \,,     \label{eq:f_0_def1} \\
 f_0^2 &= + 1/3 f_{\rm PS}^2 - 4/3 f_K^2 + f_l^2 + f_s^2 \,, \label{eq:f_0_def2} \\
 f_0^2 &= + 8/3 f_{\rm PS}^2 -16/3 f_K^2 + 3 f_s^2 \,.       \label{eq:f_0_def3}
\end{align}
It is important to note that these relations are leading order expressions in $\chi$PT. Moreover, since they are derived from continuum $\chi$PT, they differ by scaling artifacts of $\mathcal{O}(a^2)$, if applied to our lattice data. Nevertheless, this does not represent a serious drawback as the above expressions are of the same order in the chiral expansion as the Witten-Veneziano formula in Eq.~(\ref{eq:W-Vformula}) itself. In the following, we will refer to the three definitions Eqs.~(\ref{eq:f_0_def1}--\ref{eq:f_0_def3}) of $f_0$ as D1, D2 and D3, respectively.

\subsection{Lattice calculation}
For the computation of $f_0$ from the mixing parameters in the $\eta$,$\eta'$--system, we consider pseudoscalar operators in the physical basis:
\begin{equation}
 \mathcal{P}_l^{0,phys} = \frac{1}{\sqrt{2}} \bar{\psi}_l i \g{5} \psi_l \,, \quad \mathcal{P}^{\pm,phys}_h = \bar{\psi}_h i \g{5} \frac{1\pm\tau^3}{2}\psi_h \,,
\end{equation}
where we have introduced a degenerate light doublet $\psi_l$ and a non-degenerate heavy-doublet $\psi_h$ as required for the twisted mass formulation. However, in the actual calculation of the mixing parameters we drop the physical charm component.

In the twisted basis we again need the ratio $Z_P/Z_S$ due to the mixing of scalar and pseudoscalar currents in the heavy quark sector. In Tab.~\ref{tab:tabNf211} we give the values for $Z_P/Z_S$ determined from two different methods denoted by M1 and M2 (c.f. Ref.~\cite{Carrasco:2014cwa}) and the values of $r_0/a$ at each value of $\beta$, corresponding to the three lattice spacings.
\begin{table}
 \centering
 \begin{tabular}{@{\extracolsep{\fill}}ccccc}
  \hline\hline
  $\beta$ & $r_0/a$ & $a$ [fm] & $Z_P/Z_S$ (M1) & $Z_P/Z_S$ (M2) \\
  \hline\hline
  1.90 & 5.31(8) & 0.0885(36) & 0.699(13) & 0.651(6) \\
  1.95 & 5.77(6) & 0.0815(3)  & 0.697(7)  & 0.666(4) \\
  2.10 & 7.60(8) & 0.0619(18) & 0.740(5)  & 0.727(3) \\
  \hline\hline
 \end{tabular}
 \caption{Chirally extrapolated values of $r_0/a$ and $Z_P/Z_S$ from the two methods M1 and M2 as discussed in \cite{Carrasco:2014cwa} for our dynamical simulations.} 
 \label{tab:tabNf211}
\end{table}

The amplitudes (and masses) are extracted from a $2\times2$ correlation function matrix by solving a generalized eigenvalue problem after subtracting the excited states in the connected pieces as discussed in Ref.~\cite{Michael:2013gka}. Errors are computed from bootstrapping with 1000 samples and autocorrelations are taken into account by appropriate binning. The relevant disconnected diagrams are computed using stochastic volume sources and in the light quark sector we apply the one-end trick for variance reduction \cite{Boucaud:2008xu}.

\begin{figure}[t]
 \begin{center}
  \includegraphics[width=0.45\textwidth]{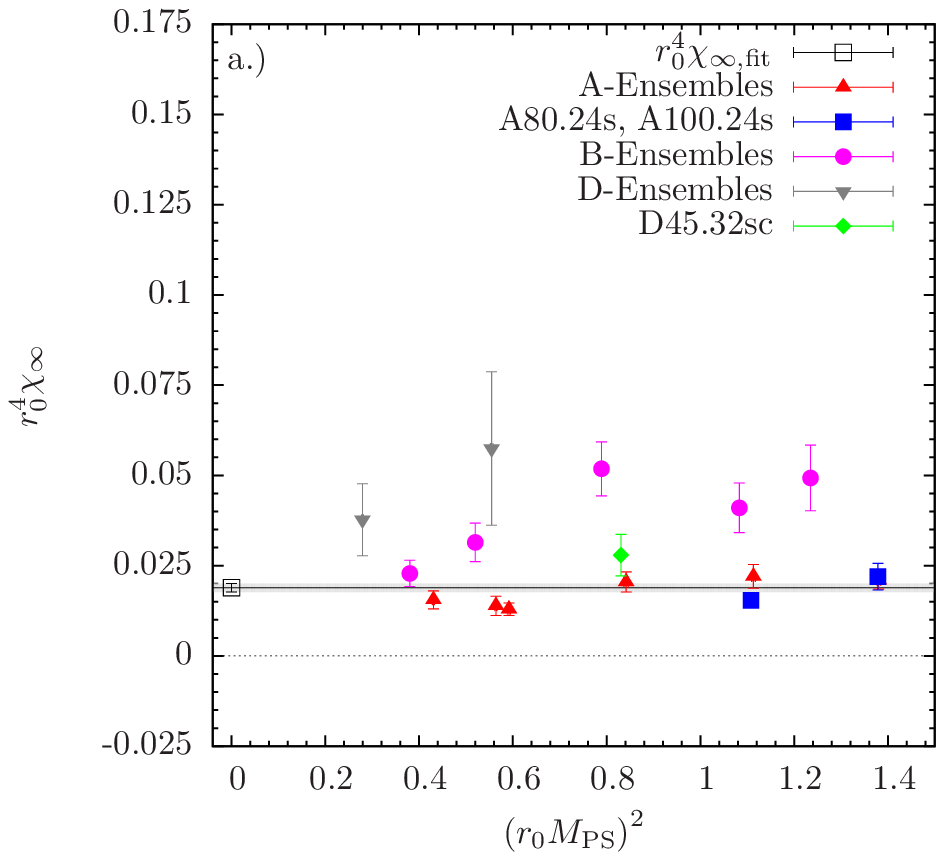}
  \includegraphics[width=0.45\textwidth]{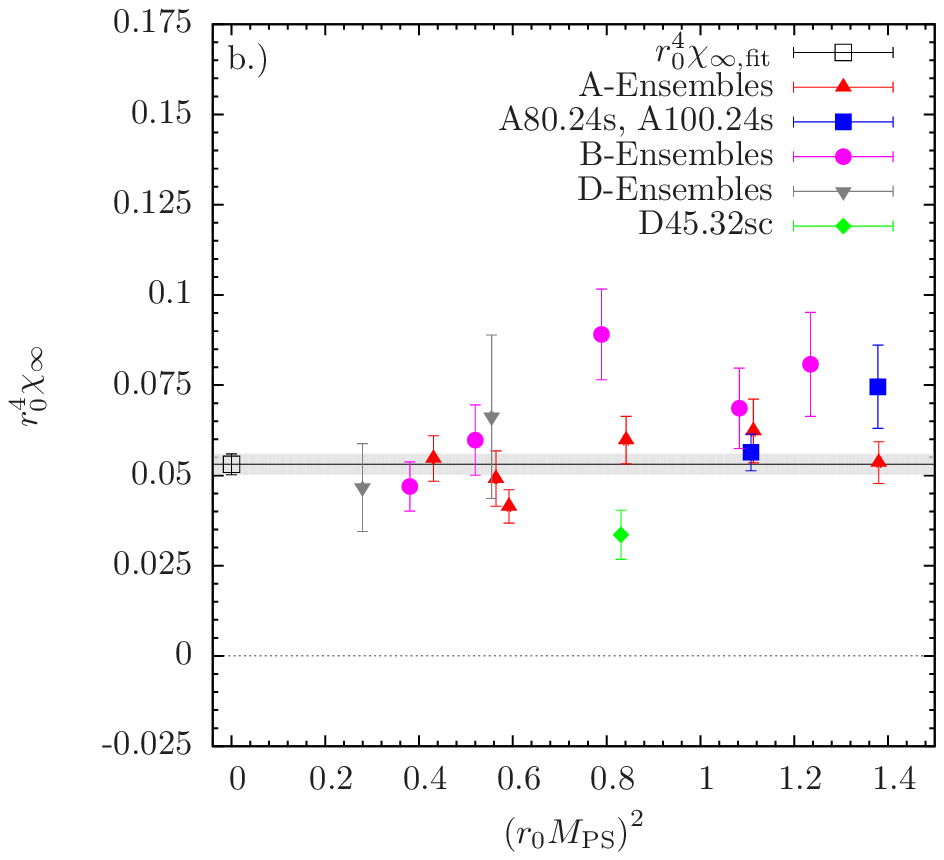}
 \end{center}
 \vspace{-0.75cm}
 \caption{Example results for $r_0^4\chi_\infty$ as a function of $(r_0 M_{\rm PS})^2$ computed from meson masses and $f_0$ from the definition in Eq.~(\protect\ref{eq:f_0_def1}). The values of $Z_P/Z_S$ used for panels a.) and b.) are taken from M1 and M2, respectively, as listed in Tab.~\protect\ref{tab:tabNf211}. Both plots show the chirally extrapolated value from a constant fit.}
 \label{fig:chi_fullQCD}
\end{figure}

In Fig.~\ref{fig:chi_fullQCD} we show two example plots for $r_0^4\chi_\infty$ calculated using definition D1 for $f_0$ and values for $Z_P/Z_S$ from method M1 and M2, respectively. Clearly, the two methods for $Z_P/Z_S$ lead to different systematic effects for the resulting values of $r_0^4\chi_\infty$ even when using the same definition for $f_0$. In addition, we find that the three definitions of $f_0$ also differ by lattice artifacts when using the same definition for $Z_P/Z_S$. The first definition D1 shows the largest lattice artifacts, as well as the most significant dependence on the choice of $Z_P/Z_S$. Applying a constant fit in $(r_0 M_\mathrm{PS})^2$ does not provide a good description of the data in this case, as can be seen from the left panel of Fig.~\ref{fig:chi_fullQCD}, leading to a $\chi^2/dof$ value of 4.90. For the right panel, the constant fit yields a better $\chi^2/dof$ value of $2.19$. In general definitions D2 and D3 lead to a reasonable agreement with a constant extrapolation in $(r_0M_\mathrm{PS})^2$ with $\chi^2/dof \lesssim 2$.

\section{Comparison of results}
\begin{figure}[t]
  \centering
  \includegraphics[width=0.5\textwidth]{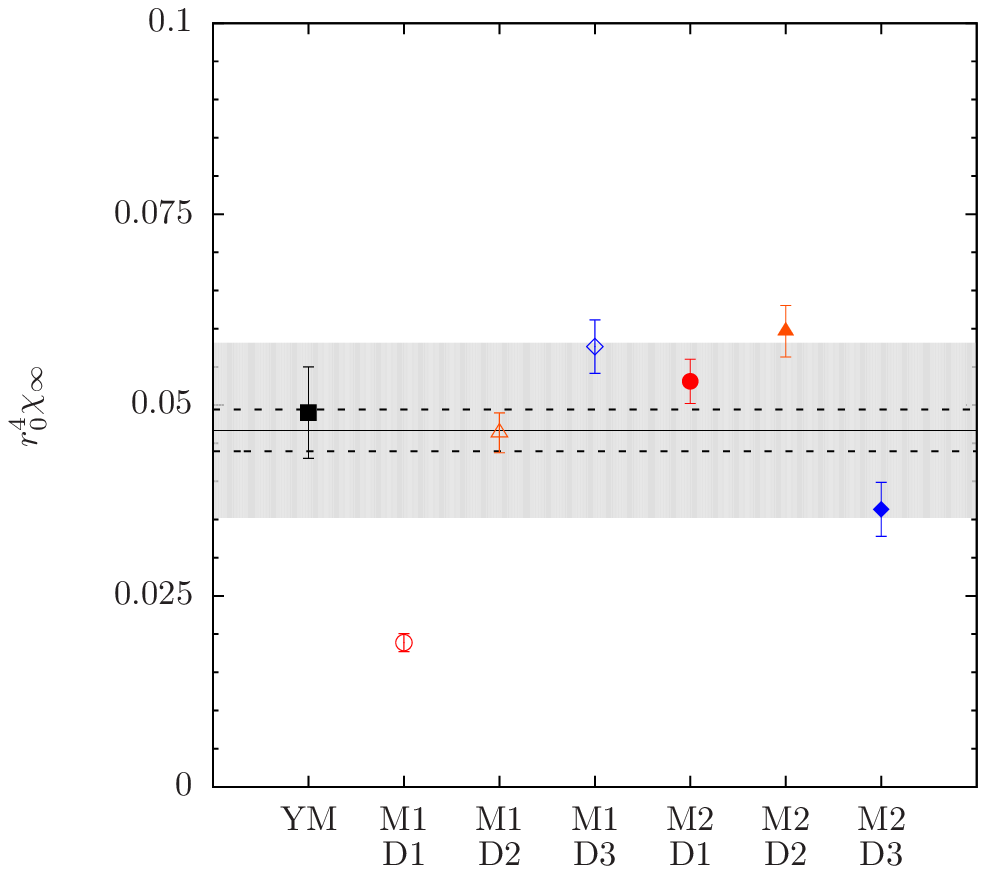}
  \caption{Results from pure Yang-Mills theory (YM) and dynamical simulations. Open and closed symbols correspond to the two sets of values for $Z_P/Z_S$ (M1, M2). The three definitions D1, D2 and D3 for $f_0$ are indicated by circle, triangle and diamond symbols, respectively. The solid black line represent the final, $p$--value weighted average from dynamical simulations and its statistical error is given by the dotted lines. The gray band represents its systematic error; see text.}
  \label{fig:comparison}
\end{figure}

For our final result of $\chi_\infty$ from the dynamical simulations, we weight each fit by $w = 1 - 2|p - 0.5| \,,$ where $p$ denotes the corresponding $p$--value and take the average over all fits, leading to:
\begin{equation}
  r_0^4\chi_\infty=0.047(3)_\stat(11)_\sys\,. \nonumber
\end{equation}
The systematic error is given by the mean absolute deviation from the central value and should reflect the uncertainties from residual cutoff and strange quark mass effects. In Fig.~\ref{fig:comparison} we show a compilation of the results from a constant fit for all definitions of $f_0$ and the two methods for $Z_P/Z_S$. Within errors we find good agreement with the result from pure Yang-Mills theory, which is also included in the plot.

Another possibility to deal with the residual effects of quark mass dependence and the lattice spacing is to include additional, higher order terms in the fit function
\begin{equation}
 f\l(r_0^4 \chi_\infty, (r_0 M_\mathrm{PS})^2, (r_0 M_K)^2, (a/r_0)^2\r) = r_0^4\chi_\infty + c_1 (r_0 M_\mathrm{PS})^2 + c_2 (r_0 M_K)^2+ c_3 (a/r_0)^2\,,
 \label{eq:lin_fit}
\end{equation}
where $c_i$ with $i=1,2,3$ denotes the free fit parameters. This fit model leads to improved $\chi^2/\mathrm{dof}$ values of $\mathcal{O}(1)$ with the exception of the data point M1D1, which is still worse. Most of the additional terms are poorly constrained by the data and are close or compatible with zero. In general, this fit ansatz leads to much larger statistical errors and all results for $r_0^4 \chi_\infty$ are compatible within errors. Taking the average of all six fits weighted by their respective p-value and statistical errors yields $r_0^4\chi_\infty=0.051(24)_\stat$, in agreement with the result from constant fits.

\acknowledgments{The computer time for this project was made available to us by the John von Neumann-Institute for Computing (NIC) on the JUDGE and Jugene systems in J{\"u}lich and the IDRIS (CNRS) computing center in Orsay. Further computational resources were provided by SuperMUC at LRZ in Garching and the PC cluster in Zeuthen.}


\newpage
\begin{thebibliography}{1}
\bibitem{Witten:1979vv}
  E.~Witten,
  Nucl.\ Phys.\ B {\bf 156} (1979) 269.

\bibitem{Veneziano:1979ec}
  G.~Veneziano,
  Nucl.\ Phys.\ B {\bf 159} (1979) 213.

\bibitem{Feldmann:1999uf}
  T.~Feldmann,
  Int.\ J.\ Mod.\ Phys.\ A {\bf 15} (2000) 159
  [hep-ph/9907491].

\bibitem{Cichy:2015jra}
  K.~Cichy {\it et al.} [ETM Collaboration],
  JHEP {\bf 1509} (2015) 020
  [arXiv:1504.07954 [hep-lat]].

\bibitem{Iwasaki:1985we}
  Y.~Iwasaki,
  Nucl.\ Phys.\ B {\bf 258} (1985) 141.

\bibitem{Jansen:2009xp}
  K.~Jansen and C.~Urbach,
  Comput.\ Phys.\ Commun.\  {\bf 180} (2009) 2717
  [arXiv:0905.3331 [hep-lat]].

\bibitem{Jansen:2005gf}
  K.~Jansen {\it et al.} [XLF Collaboration],
  Phys.\ Lett.\ B {\bf 619} (2005) 184
  [hep-lat/0503031].

\bibitem{LuscherGiusti}
L.~Giusti and M.~L{\"u}scher,
\newblock JHEP {\bf 0903}, 013 (2009).

\bibitem{Luscher:2010ik}
  M.~Luscher and F.~Palombi,
  JHEP {\bf 1009} (2010) 110
  [arXiv:1008.0732 [hep-lat]].

\bibitem{Cichy:2013rra}
  K.~Cichy {\it et al.} [ETM Collaboration],
  JHEP {\bf 1402} (2014) 119
  [arXiv:1312.5161 [hep-lat]].

\bibitem{Baron:2010bv}
  R.~Baron {\it et al.},
  JHEP {\bf 1006} (2010) 111
  [arXiv:1004.5284 [hep-lat]].

\bibitem{Cichy:2014yca}
  K.~Cichy, E.~Garcia-Ramos and K.~Jansen,
  JHEP {\bf 1504} (2015) 048
  [arXiv:1412.0456 [hep-lat]].

\bibitem{Frezzotti:2000nk}
  R.~Frezzotti {\it et al.} [Alpha Collaboration],
  JHEP {\bf 0108} (2001) 058
  [hep-lat/0101001].

\bibitem{Ottnad:2012fv}
  K.~Ottnad {\it et al.} [ETM Collaboration],
  JHEP {\bf 1211} (2012) 048
  [arXiv:1206.6719 [hep-lat]].

\bibitem{Michael:2013gka}
  C.~Michael {\it et al.} [ETM Collaboration],
  Phys.\ Rev.\ Lett.\  {\bf 111} (2013) 18,  181602
  [arXiv:1310.1207 [hep-lat]].

\bibitem{Carrasco:2014cwa}
  N.~Carrasco {\it et al.} [ETM Collaboration],
  Nucl.\ Phys.\ B {\bf 887} (2014) 19
  [arXiv:1403.4504 [hep-lat]].

\bibitem{Boucaud:2008xu}
  P.~Boucaud {\it et al.} [ETM Collaboration],
  Comput.\ Phys.\ Commun.\  {\bf 179} (2008) 695
  [arXiv:0803.0224 [hep-lat]].
\end{thebibliography}
\end{document}